# Slow diffusion and long lifetime in metal halide perovskites for photovoltaics

Adrien Bercegol[1,2], F. Javier Ramos[1,3], Amelle Rebai[1], Thomas Guillemot[1,4], Daniel Ory[1,2], Jean Rousset[1,2] Laurent Lombez[1,3]

[1] IPVF, Ile-de-France Photovoltaic Institute, 30 RD 128, 91120 PALAISEAU, France

[2] EDF R&D, 30 RD 128, 91120 PALAISEAU, France

[3] CNRS, Ile-de-France Photovoltaic Institute UMR 9006, 30 RD 128, 91120 PALAISEAU, France

[4] Licorne Laboratory, ECE Paris, 37 Quai de Grenelle, 75015 Paris, France

## Abstract

Metal halide perovskites feature excellent absorption, emission and charge carrier transport properties. These materials are therefore very well suited for photovoltaics applications where there is a growing interest. Still, questions arise when looking at the unusual long carrier lifetime that, regarding the micrometric diffusion length, would imply a very low diffusion coefficient as compared to commonly used photovoltaic absorbers. In this paper, we provide an experimental insight into this long lifetime of charge carriers in slow-motion. Our approach relies on an improved model to analyze time-resolved photoluminescence decays at multiple fluence levels that includes charge carrier transport, photon recycling, and traps dynamics. The model is verified on different interface properties. Moreover, we investigate various perovskite absorbers such as mixed alloys with cesium content. For most of the perovskite based materials we analyzed, the band-to-band recombination rate remains close to the radiative limit, leading to the expected sub-microsecond lifetime. The slow diffusion of charge carriers is observed with values of the diffusion coefficient $D$ around $5\times10^{-3}$ cm$^2$/s. Nonetheless, the power conversion efficiency remains high. The observations might be related to the debated coexistence of a direct and indirect bandgap slowing down the recombination process.

## Introduction

The interest of the photovoltaic community for the perovskite solar cells has been burgeoning in the past years, as their power conversion efficiency (PCE) rose from 3% to 22.7%[1–3]. Various application fields arise from their widely tunable bandgap[4], all of them benefiting from the high diffusion length in perovskite layers[5], their excellent absorption properties[6] and their high defect tolerance[7]. In this paper, we focus on their diffusion properties, as they play a crucial role in solar cell operation, during which charge carriers diffuse to the extraction layers[8]. These diffusion properties have two complementary aspects: the sensitivity to a carrier concentration gradient (diffusion coefficient $D$), and the average time available for the diffusion process (lifetime $\tau$). A wide range of $D$ values have been reported in the literature[4,9–11]. However, a recent model based on the co-existence of a direct & indirect bandgap[12–14] suggests that a long lifetime would compensate a slow diffusion process inside highly efficient perovskite solar cells[15]. Here we propose to evaluate $D$ and $\tau$ by measuring Time-Resolved Photoluminescence (TRPL) decays in perovskite layers.

Time resolved optoelectronic characterization techniques have thoroughly been used to investigate the transport properties inside perovskite solar cells. At the picosecond scale[16], transient optical transmission spectroscopy allowed determining the best electron/hole transport layers (E/HTL), by studying the charge carrier extraction speed at interfaces[17,18]. At the ns scale, bulk transport and recombination become dominant[16]. TRPL decays are well adapted to this time scale, and their analysis paved the way towards a deeper understanding of charge carrier transport inside perovskite absorbers[19,20]. At first, the low binding energy of the exciton[5,21] explained their negligible concentration under realistic excitation conditions (1 sun illumination, room temperature), which initiated the switch from an organic point of view to an inorganic one[11,16]. In a second phase, this technique also allowed to underline the predominant role of trap states in recombination kinetics, for various chemical compositions of perovskite[22–25]. Considering the multiple transient phenomena in bulk perovskite, the simple PL lifetime is not sufficient to assess its quality[26], on the contrary to physical constants extracted thanks to more detailed models.

In this work, we start from a well-established model including traps[22,27], which we complete by considering the in-depth diffusion of charge carriers. Its impact had been predicted for TRPL in thin films[28]. A vertical geometrical component is added to the model, which ultimately leads us to the recombination velocity both at the front and back interface. It is noteworthy that we acquired TRPL decays following a wide-field pulsed illumination, which avoids the artefacts due to lateral diffusion. This allows us to focus even more precisely on the in-depth diffusion process in various perovskite absorbers and compare them to emerge common properties.

While the general formula for hybrid organic-inorganic halide perovskites is fixed by the crystalline structure they share ($ABX_3$, see Suppl. Fig. S1C), numerous possibilities in the chemical composition arise for the choice of the cation $A^+$ (methylammonium cation $MA^+$, formamidinium cation $FA^+$, $Cs^+$, $Rb^+$)[4,29–31], the halide $X^-$ ($I^-$; $Br^-$; $Cl^-$) [4,18,32], and the central metal cation $B^{2+}$ ($Pb^{2+}$ or $Sn^{2+}$)[33]. For our study, we selected various absorber layers for their bandgap adapted to photovoltaic (PV) applications, and for their relative stability to phase segregation[34,35]. The number of monovalent cations $A^+$ varies from one to three. The simplest hybrid perovskite compound is $MAPbI_3$ (noted $P_1$), to which $FA^+$ ($MA_{0.83}FA_{0.17}Pb(I_{0.83}Br_{0.17})_3$) (noted $P_2$), $FA^+$ and $Cs^+$ (($MA_{0.83}FA_{0.17})_{0.95}Cs_{0.05}Pb(I_{0.83}Br_{0.17})_3$) (noted $P_3$) or $FA^+$ and $Rb^+$ (($MA_{0.83}FA_{0.17})_{0.95}Rb_{0.05}(I_{0.83}Br_{0.17})_3$) (noted $P_4$) are added.

For TRPL characterization, samples were prepared with an architecture both perovskite/fluorine-doped tin oxide (FTO) and perovskite/mesoporous(mp)-$TiO_2$/blocking(bl)-$TiO_2$/FTO, which is represented in Figure 1-B. FTO was commercially available while bl-$TiO_2$ was synthetized by spray pyrolysis and mp-$TiO_2$ was spun coated and subsequently sintered. Perovskites were prepared with minor modifications regarding previously reported procedures (Ahn *et al.*[36] for $P_1$, and Saliba *et al.*[29,31] for $P_2$, $P_3$, and $P_4$). Concerning the devices used for PV characterization, Spiro-OMeTAD was deposited by spin coating, as well as Au by thermal evaporation over $TiO_2$ containing devices to complete the solar cell. More details concerning the sample preparation is available in the Supplementary Information.

## PV characterization

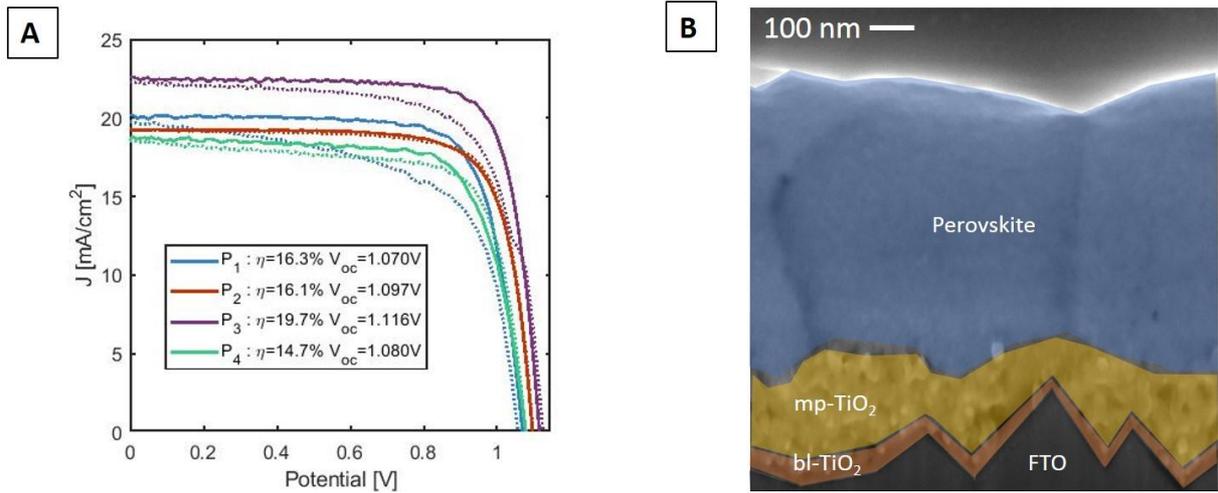

*Figure 1 A-J-V characteristics recorded in reverse conditions (80mV/s, full line) and forward conditions (80 mV/s, dotted line) of the devices based on perovskite absorber with various chemical composition. Extracted electrical parameters are indicated in the caption. B- Focused Ion Beam (FIB) image, highlighting the layer structure of the sample probed by TRPL, when mp/bl-$TiO_2$ is the back interface..*

The cell performances are strongly linked to the absorber composition as shown in the Figure 1-A, where the electrical parameters from our different record cells fabricated in our lab are summarized. The PCE of the mono-cation perovskite $P_1$ reaches a maximum value of η = 16.3 %. However, the voltage scan direction has a drastic influence on the PCE, as the forward scan (increasing potential) leads to an absolute 3 % loss in PCE (η = 13.1 %).

This hysteresis is largely mitigated by the partial substitution of MA through an additionial organic cation (FA) in the perovskite structure. If the maximum PCE remains closely the same (η = 16.1 % for $P_2$), the sensitivity of the electrical parameters to the measurement conditions is reduced. The PCE in forward conditions is 16.2 % and the currents of the EQE and the I-V closely match. Interestingly, the cell performance is further enhanced with the incorporation of Cs cation ($P_3$), as its PCE reaches 19.7 % (η = 17.6 % for a forward scan. On the contrary, the use of the Rb as third cation in $P_4$ leads to a lower cell performance with a maximum PCE of 14.7 %.

## Time-Resolved Photoluminescence

### Set-up

The main components of the TRPL set-up are an intensified, time-gated, electron-multiplied CCD (em-ICCD) camera (PIMAX4, Princeton instruments) and a pulsed laser (TALISKER, Coherent). The latter delivers 15 ps-wide pulses (λ=532 nm) at a repetition rate $f_0$=40 kHz, and triggers the em-ICCD camera. For our study, the gating time was varied with a 2ns-step between $t=t_{pulse}$ (taken as reference) and $t=1$ μs, whereas the gate width remained constant at 2 ns. The wide-field illumination is obtained thanks to a home-built opto-mechanical set-up that also filters laser speckles[37]. The intensity of the photon flux incident on the sample was varied, over three order of magnitude ($\Phi_{0,1}$=2.5x10$^{10}$; $\Phi_{0,2}$=2.25x10$^{11}$; $\Phi_{0,3}$=1.6x10$^{12}$/pulse/cm$^2$). The illuminated area (≈1 mm$^2$) is larger than the light collection area (≈0.01 mm$^2$), itself significantly larger than the diffusion area inside the samples (≈1-10 μm$^2$). Hence, lateral diffusion of charge carriers does not lead to any artefacts[38] . The probed PL signal originates from the perovskite layer, as shown by its spectrum (see Figure S2). We studied two neighboring regions of the same sample (100µm×10µm), where PL is spatially averaged. They correspond to two cell architectures: perovskite/FTO and perovskite/mp-TiO$_2$/bl-TiO$_2$/FTO, as sketched in Figure S1D.

Model

To model the charge carrier kinetics in the perovskite layer, we consider 3 distinct charge carrier populations: the electrons excited in the conduction band $n$, the holes in the valence band $p$ and the trapped electrons $n_T$, which concentration has to remain lower than the total traps concentration $N_T$. These populations are represented on an energetic diagram in Figure 2-A, where arrows indicate the authorized transitions along with their associated rates $R_{pop}$ (trap capture), $R_{dep}$ (trap-assisted recombination) and $R_{eh}*$ (net band-to-band recombination, including generation due to photon recycling).

Here, it is useful to focus on the evolution of the trapped population $n_T$ in a pulsed excitation regime. Its time average value stabilizes at $n_{T,0}$[39], as the relaxation lifetime have typical values within the µs to ms range[22,35]. Henceforth, it can be assumed constant during a TRPL decay, which lasts at most few µs. More details about $n_{T,0}$ calculations are available in the Supplementary Information.

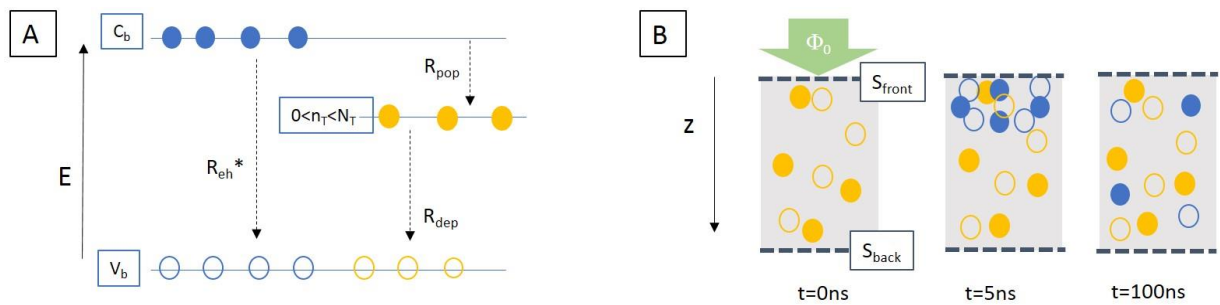

Figure 2 A- Sketch showing the allowed energetic transitions in perovskite. $C_b$ and $V_b$ stand for conduction and valence band. $R_{eh*}$ is the net band-to-band recombination rate, $R_{pop}$ is the trap capture rate, $R_{dep}$ is the trap-assisted recombination rate. $n_T$ is the trapped electron concentration, which has to remain lower than the total trap concentration $N_T$. B- Sketch showing the slow in-depth diffusion of charge carriers, following an incoming photon flux $\Phi_0$ at t=0ns. Full (empty) circles stand for electrons (holes). Blue symbols stand for photo-generated charge carriers, while yellow ones stand for trapped electrons and photo-doped holes.

In addition to the charge carrier recombination kinetics, one should also consider the in-depth transport. In fact, the absorption length of the perovskite layer ($1/\alpha_{532}$ = 100 nm)[40] is significantly lower than its thickness ($z_0$ = 500 nm), and is not significantly affected by chemical compositional tailoring at the excitation wavelength (λ=532nm)[41–44]. As a consequence, the photo-generated carriers are concentrated close to the front interface at short times after the laser pulse, as illustrated by the sketch in Figure 2-B and by numerical simulations displayed in Figure S3. The time required for the vertical homogenization mainly depends on $D$. In fast diffusion materials such as III-V semi-conductors, this effect has no influence on TRPL signal as the carriers homogenization in-depth occurs in a few ps[45]. It becomes visible in some photovoltaic absorbers such as copper-indium-gallium-selenide[28] and as we will see, it has a significant effect on TRPL decays in perovskite materials. Therefore, we assume that the two physical phenomena sketched in Figure 2-B dominate the charge carriers kinetics in a pulsed excitation regime. We also assume a local electro-neutrality:

$$p(z,t) = n_{T,0} + n(z,t) \quad (E1)$$

This implicitly induces that the diffusion coefficients for electrons and holes are similar, which is consistent with previous observations in intrinsic perovskite material[8]. This also allows us to reproduce the whole transport of charge carriers by writing a simple rate equation for a single population (E2), with spatial Von Neumann boundary conditions (E3) and (E4):

$$\frac{\partial n}{\partial t} = -R_{eh}^* n(n_{T,0} + n) - R_{pop} n(N_T - n_{T,0}) + D_n \frac{\partial^2 n}{\partial z^2} \quad (E2)$$

$$\forall t, z = 0 \rightarrow D_n \frac{\partial n}{\partial z} = S_{front} n \quad (E3)$$

$$\forall t, z = z_0 \rightarrow D_n \frac{\partial n}{\partial z} = -S_{back} n \quad (E4)$$

The temporal initial condition is determined by the absorption properties of the sample, before any in-depth diffusion or front surface recombination can happen:

$$t = 0 \rightarrow n = \Phi_0 \alpha_{532} \exp(-\alpha_{532} z) \quad (E5)$$

Once this differential system solved, the PL intensity $I_{PL}$ remains to be calculated. For the considered intrinsic semi-conductor, it can be expressed as:

$$I_{PL} \propto \int_0^{z_0} R_{eh}^* n(n_{T,0} + n) \exp(-\alpha_{PL} z) dz \quad (E6)$$

In this formula, partial reabsorption of the PL signal is considered thanks to the coefficient $\alpha_{PL}$, which corresponds to the average absorption coefficient for the photoluminescence signal (derived with (ES7) in the Supplementary Material). This PL reabsorption notably induces a photon recycling phenomenon inside the absorber, which we take into account with a relative correction of the fitted external radiative coefficient $R_{eh}^*$, in order to determine the internal radiative coefficient $R_{eh}=R_{eh}^*/(p_{escape})$, with $p_{escape}$=12.5% [46–48,27]. Eventually, an optimization algorithm is employed to minimize the logarithmic difference from experimental to numerical transients. The robustness of our method is notably enhanced by the simultaneous fitting of transients obtained at various fluence levels.

Classical optoelectronic properties such as the lifetime $\tau_n$ and the diffusion length $L_n$ can be derived from this transport model thanks to the following procedure. If one starts from the simple formula $L_n=\sqrt{(D_n\tau_n)}$, a difficulty quickly arises as $\tau_n$ is strongly injection-dependent for an intrinsic semi-conductor with a bimolecular recombination regime. Here, we determine the charge carrier concentration by solving the differential system (ES1-3) under a steady state illumination, using the fitted $R_{eh}^*$ and $N_T$ coefficients. It converges to a unique stable solution, giving access to the steady state filled trap concentration $n_{T1sun}$ and electron concentration $n_{1sun}$. Interestingly, both increase significantly when $N_T$ grows, as displayed in Table S1. Thereafter, $\tau_{1sun}$ and $L_{1sun}$ can be calculated thanks to formulas (E7-8). Though trap depopulation plays a role in steady state condition, it is considered as $n_{T1sun}<N_T$ and hence only $R_{pop}$ and $R_{eh}$ appear in this formula.

$$\tau_{1sun} = \frac{n_{1sun}}{R_{1sun}} = \frac{1}{R_{pop}(N_T - n_{T,1sun}) + R_{eh}^*(n_{1sun} + n_{T,1sun})} \quad (E7)$$

$$L_{1sun} = \sqrt{D_n \tau_{1sun}} \quad (E8)$$

Results

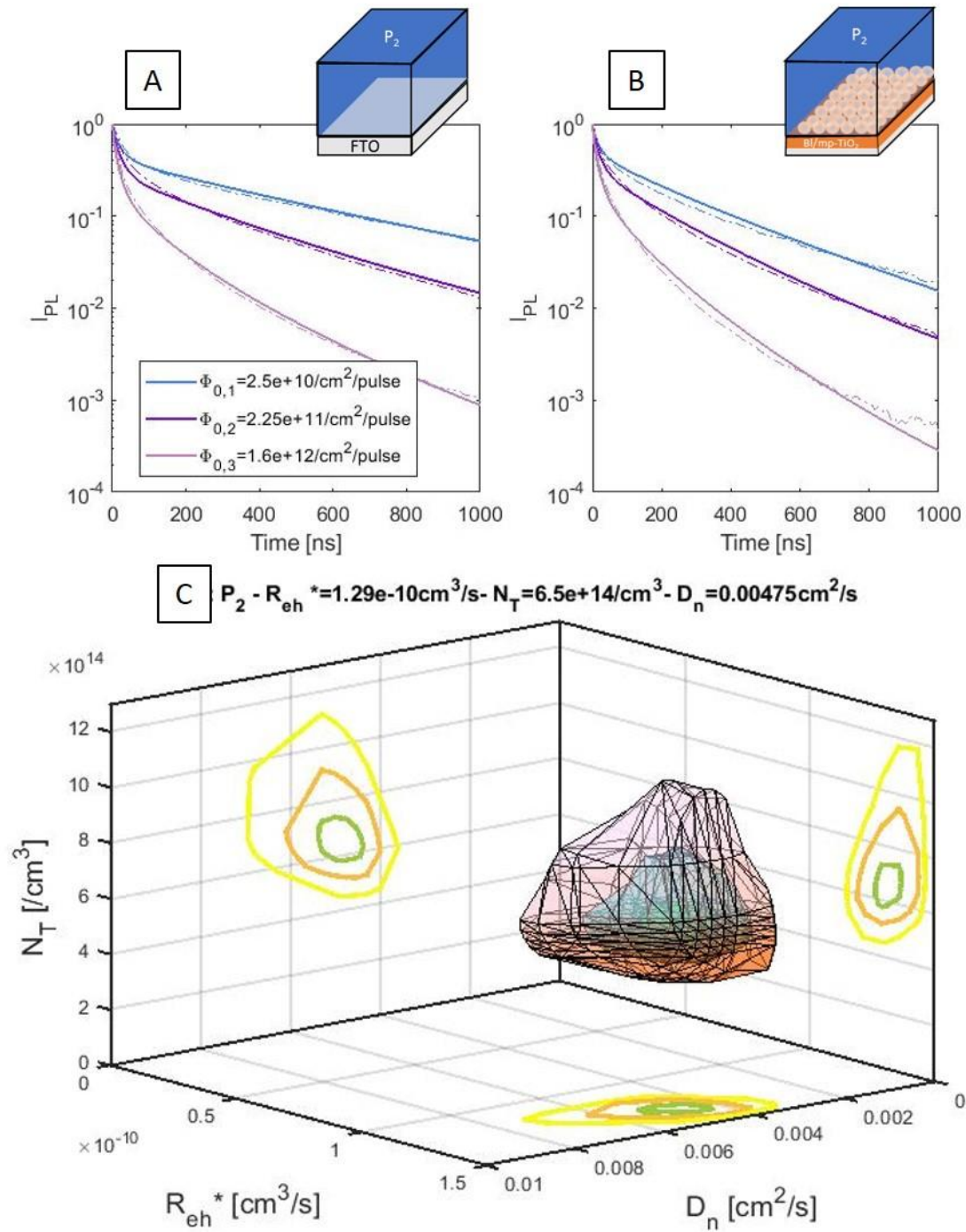

Figure 3 Experimental (dotted lines) and numerically fitted (plain lines) TRPL transients for a mixed cation perovskite $MA_{0.83}FA_{0.17}Pb(I_{0.83}Br_{0.17})_3$ (also $P_2$) layer deposited on FTO (A) or on mesoporous-$TiO_2$ (B). Various orders of magnitude of the fluence level $\Phi_0$ have been tested. (C) Iso-surfaces showing a set of parameters ($R_{eh}*$ $D_n$ $N_T$) leading to a reconstruction error not higher than 10 % (green), 20% (orange), 50% (yellow) of the optimal reconstruction error. The green-orange one is used for confidence intervals indicated in Table 1.

Our model was first tested on a mixed cation perovskite layer directly deposited on FTO. Experimental TRPL transients obtained at increasing fluence levels are displayed in Figure 3-A, along with the optimized numerical ones. In the fitting process, the trap-related recombination rates were fixed to literature values ($R_{pop}$ = 2·$10^{-10}$ $cm^{-3}s^{-1}$, $R_{dep}$ = 8·$10^{-12}$ $cm^{-3}s^{-1}$)[22], while the main optoelectronic parameters related to the perovskite layer were fitted, yielding an external radiative recombination coefficient $R_{eh}^* = 1.3 \cdot 10^{-10}$ cm$^{-3}$s$^{-1}$, a trap concentration $N_T = 6.5 \cdot 10^{14}$ cm$^{-3}$, a diffusion coefficient $D_n = 4.8 \cdot 10^{-3}$ cm$^2$s$^{-1}$, as well as a front interface recombination velocity $S_{air}$, with values ranging between 140 cm/s at high fluence and 100 cm/s at low fluence. The back interface recombination velocity was taken as reference ($S_{FTO}$ = 0 cm/s), which is justified by a control experiment realized on a mixed cation perovskite layer deposited on glass. It featured similar TRPL decays as those acquired on FTO (see Suppl. Fig S8), which validates the assumption of a weak surface recombination velocity on FTO. Moreover, the narrow confidence intervals, which are extracted from Figure 3-C and included in Table 1, witness for the precision of the fit. It is worth mentioning that the fitted parameters allow to fit the three whole decays, the excitation flux being the only variable parameters.

A complementary experiment was then realized for the same perovskite deposited on mesoporous TiO$_2$. Resulting TRPL transients are shown in Figure 3-B. Even though the decays do not look like those acquired on FTO, they can still be numerically reproduced by our model. Furthermore, they are fitted thanks to a single parameter variation, as $S_{back}$ is increased to $S_{TiO2}$ = 40 cm/s on TiO$_2$, whereas it was taken as a reference $S_{FTO}$ = 0 cm/s on FTO. This assesses the consistency of our model.

We repeated this same experiment for samples P$_1$, P$_3$ and P$_4$, having a similar architecture but different absorbers. Experimental and numerical transients are available in Figure S4 to S6 (see Supplementary Information). Again, our model allows a fair reproduction of TRPL transients in the FTO/P$_n$ architecture for all samples considered, and allows to extract key optoelectronic properties for the bulk and the perovskite/air interface. The confidence intervals indicated in Table 1 are narrow, except for the diffusion coefficient/length in MAPbI$_3$ (which we attribute to the quick degradation of this single cation perovskite). Concerning the second architecture on mesoporous TiO$_2$, the previously explained introduction of $S_{TiO2}$ leads to a successful fitting for sample P$_1$ and P$_4$, but is not sufficient to explain the physical phenomena involved at the TiO$_2$ interface in Cs-containing sample (P$_3$).

| | Sample P$_1$ MAPbI$_3$ | Sample P$_2$ MA$_{0.83}$FA$_{0.17}$... Pb(I$_{0.83}$Br$_{0.17}$)$_3$ | Sample P$_3$ (MA$_{0.83}$FA$_{0.17}$)$_{0.95}$ ...Cs$_{0.05}$... Pb(I$_{0.83}$Br$_{0.17}$)$_3$ | Sample P$_4$ (MA$_{0.83}$FA$_{0.17}$)$_{0.95}$ ...Rb$_{0.05}$... Pb(I$_{0.83}$Br$_{0.17}$)$_3$ |
|---|---|---|---|---|
| $R_{eh}^* = R_{eh}/p_{escape}$ [x10$^{-10}$ cm$^{-3}$·s$^{-1}$] | 4.0 [3.0 ; 5.0] | 1.3 [1.2 ; 1.4] | 1.2 [1.1 ; 1.3] | 1.1 [0.95 ; 1.2] |
| $D_n$ [x10$^{-3}$ cm$^2$·s$^{-1}$] | 0.8 [0.1 ; 100] | 4.7 [3.5 ; 6.5] | 6 [4.8 ; 8.0] | 3.1 [2.1 ; 4.8] |
| $N_T$ [x10$^{14}$ cm$^{-3}$] | 560 [500 ; 600] | 6.5 [5.5 ; 9] | 4.2 [3.5 ; 5.5] | 80 [75 ; 85] |
| $L_{1sun}$ [x10$^{-4}$ cm] | 0.06 [0.006 ; 0.75] | 0.66 [0.52 ; 0.79] | 0.78 [0.67 ; .93] | 0.37 [0.3 ; 0.5] |
| $\tau_{1sun}$ [ns] | 43 [38 ; 55] | 886 [780 ; 960] | 1009 [930 ; 1100] | 440 [420 ; 520] |
| $S_{air}$ [cm·s$^{-1}$] $\Phi_{0,1}/\Phi_{0,2}/\Phi_{0,3}$ | 120/180/180 | 100/100/140 | 60/80/140 | 60/75/100 |

| $S_{TiO2}$ [cm·s$^{-1}$] $\Phi_{0,1}/\Phi_{0,2}/\Phi_{0,3}$ | 80/80/80 | 60/40/40 | Unconverged | 100/100/60 |
|---|---|---|---|---|

Table 1 Physical parameters for each sample probed. $R_{eh}$, $D_n$ $N_T$ and $S_{air}$ (perovskite/air interface) are determined by fitting TRPL transients on FTO, while $S_{TiO2}$ is determined on TiO$_2$ and corresponds to the perovskite/mp-TiO$_2$ interface. For the whole range of fitting parameters inside the confidence intervals, the reconstruction error rises by less than 20%. Diffusion length $L_{1sun}$, lifetime $\tau_{1sun}$ and associated tolerance intervals are calculated using (E7-8)

## Discussion

Thanks to a complete model for TRPL decays, we could highlight the main phenomena governing the charge carriers transport and recombination in perovskite, leading to a versatile description of their kinetics for a wide range of fluence level. The main recombination channel for charge carriers, which dictates their decay rate, changes during the TRPL decay, and also as the excitation fluence is varied. This is displayed in Table S3, where we can see that trap-assisted recombinations dominate the second part of the decay for samples with $N_T>5\times10^{15}$/cm$^3$, while the slow in-depth diffusion must be considered to explain the faster PL decay at short times, especially in low-mobility perovskite absorbers. A first justification relies on the fact that charge carriers remain concentrated in a smaller volume for a longer time, which accelerates the radiative recombination process, as electrons and hole "meet" more often[28]. Thereafter, they diffuse away from the defects located at the front interface, which eventually slows down the early decay.

On the one hand, previous physical models for perovskites predicted mono-exponential decays at low fluence[22,47]. On the other hand, other empirical models might have been developed to explain non-exponential decays, like the stretched exponential model[33,49] or the bi-exponential model[50,51]. However, they fail to reproduce transients from a power study, and can only yield relative quantification of recombination and transport processes, as illustrated by the injection-dependent "lifetimes" we determined and displayed in Table S2 if one uses a bi-exponential decay model. Here we attribute the fast first decay to the diffusion properties while keeping the second slower decay to bulk and interface defects dynamics at any fluence.

Moreover, our model allows us to quantify the recombination rate attributed to the interfaces, generally guilty of the most radiative losses in perovskite solar cells[52]. While $S_{air}$ refers to the perovskite/air interface, $S_{TiO2}$ refers to the one located at mp-TiO$_2$/perovskite. All samples considered in this study were half-cells and hence these velocities correspond to recombination at interfacial defects rather than electrical injection, which is often observed in full devices at short-circuit[4]. The interface formed at FTO/perovskite was taken as reference for $S_{back}$, as our experiment cannot effectively distinguish between bulk and back surface recombination, and we only illuminated from the air/perovskite side. In the future, one could consider acquiring two set of TRPL decays by illuminating one side after another, in order to decorrelate the influence of both interfaces[53].

Still, the successful fitting on Z$_2$ allows us to assess that the mp-TiO$_2$/perovskite induces a higher recombination current than the FTO/perovskite one. This can be explained by the larger contact surface at such a meso-structured interface. Now, $S_{TiO2}$ remains lower than $S_{air}$, showing a preserved interface crystalline quality, in comparison to the one obtained after exposure to the ambient atmosphere. Eventually, it can be noted that both $S_{air}$ and $S_{TiO2}$ demonstrate a slight injection-dependence. Even if a fine modeling of interface recombinations is out of the scope of this study, it can be noted that these recombination velocities quantify the interface trap density, which is not sufficient to describe the interface recombination current, as the Fermi level position at the interface and the eventual trap saturation possibly matter[54,55].

Concerning the diffusion coefficient, we found low values of $D_n$ being in the same order of magnitude for every sample. It implies a relative long time to obtain a homogeneous in-depth carrier distribution. A direct application of the Einstein law ($D_n=\mu_n kT/q$) leads to $\mu_n$ values comprised between 0.1 and 0.3 cm$^2$/(V.s), which locates our batch in the lower values for polycrystalline solution processed perovskites[9–11]. This might seem contradictory with the value of PCE as high as 19% obtained with such an absorber. However, these low $D_n$ values are compensated with the very long lifetime $\tau_{1sun}$ of charge carriers, which we display in Table 1. They were derived by solving our transport model under steady-state solar illumination, using the fitted $R_{eh*}$ and $N_T$ parameters (see equations (E7) & (E8)). Their high values ensure that the diffusion length $L_{1sun}$ remains larger than $z_0$, thus preserving an efficient charge carrier collection.

This experimental insight into slow diffusion could bring crucial information regarding the recent discovery of a Rashba splitting in MAPbI$_3$ [14,56,57], as well as in CsPbI$_3$ [58]. This effect linked to spin-orbit coupling leads to the coexistence of a direct and an indirect bandgap, the first one being beneficial to absorption, and the latter drastically enhancing the charge carriers lifetime. The main recombination pathway for photo-generated carriers relies then on a phonon coupling, or on the overcoming of an activation barrier[12,13]. Regarding a recent study from Kircharz et al., a slow diffusion together with high PCE is possible thanks to this particular band structure[15]. However, it should be noted that the significance of this spin-orbit coupling is still under debate in the perovskite community[48].

The trap concentration is highly dependent on the chemical composition of the considered sample, with values varying between 4.2 x10$^{14}$ cm$^{-3}$ for sample P$_3$ and 5x10$^{16}$ cm$^{-3}$ for sample P$_1$. The latter is consistent with $N_T$ values previously reported in MAPbI$_3$ absorbers, for various energetic positions[22,49,59]. The influence of traps on the charge carrier kinetics in a pulsed excitation regime mainly happens through the photo-doped charges they fix in the valence band by trapping electrons on the long term[22,24,27]. This is mainly described by their concentration N$_T$, which lower value in P$_2$ and P$_3$ confirm the effective trap passivation through cation substitution previously observed[60,61]. What's more, the good correlation between the effect of chemical composition on $N_T$ and $V_{oc}$ values underlines the significant role of these traps in solar cell operation. On the contrary to $N_T$ values, $R_{eh}*$ coefficients are quite constant and stick to low values for every sample, and photon recycling corrected values elevate up to 10$^{-9}$/cm$^3$/s, notably close to the expected radiative recombination rate[16]. This confirms the negligible role of deep defect, as the non-radiative pathways they generate would be included in this coefficient, too. Our study confirms that traps constitute the main non-radiative recombination pathway in a wide variety of bulk perovskite. If they were absent in a device with perfect interfaces, the electric open-circuit voltage would equal the radiative one, generally evaluated around 1.36V[62].

## Conclusion

In this paper, we developed a charge carrier transport model incorporating for the first time a combination of the well-established traps dynamics and the slow in-depth diffusion after a pulsed excitation. We applied it to the simultaneous fitting of TRPL decays at various fluence levels, for different PV-compatible perovskite compositions. The variation in chemical composition mainly impacts the trap concentration, while the slow diffusion remains characteristic for the whole batch. The best PCE (19.7%) obtained after Cs incorporation could be explained by a reduction in trap state density ($N_T = 4.2 \times 10^{14}/cm^3$). It is notably high despite the underlined slow diffusion, probably thanks to the coexistence of a direct and indirect bandgap, or excitonic effect, where these origins are still under discussion. In any case, this model includes a vertical component, which opens the way for independent characterization of ETL/- and HTL/perovskite interfaces.

## Acknowledgements

This project has been supported by the French Government in the frame of the program of investment for the future (Programme d'Investissement d'Avenir - ANR-IEED-002-01). Authors acknowledge Thomas Bidaud for providing SEM pictures used during the review process.

# Supplementary Information

## Device Fabrication

Fluorine-doped tin oxide (FTO)-coated glass (Solems) was etched to pattern appropriately using Zn powder and HCl (4M). Substrates were cleaned by ultrasonication in RBS detergent solution (2% in weight), rinsed with deionized water and subsequently ultrasonicated in absolute ethanol and dried. Then, the samples were burnt at 500ºC to eliminate any organic residue. At the center of the substrate, a $TiO_2$ dense buffer layer (bl-$TiO_2$) was deposited by spray pyrolysis at 450ºC using a solution containing 0.6mL of titanium diisopropoxide bis(acetyl acetonate) (75% in 2-propanol, Sigma Aldrich), 0.4mL of acetyl acetone (Sigma Aldrich) in 9mL of absolute ethanol and oxygen as carrier gas. A mesoporous $TiO_2$ (mp-$TiO_2$) layer ~150nm was deposited by spin-coating a solution of commercial 30NR-D (Dyesol) diluted in absolute ethanol (1:7 in wt. respectively). Afterward, to prepare the crystalline $TiO_2$ mesoporous particles, samples were annealed at 125ºC for 5min, 325ºC for 5min, 375ºC for 5 min, 450ºC for 15min and 500ºC for 30 min. A 1cm-wide zone was left without bilayer-$TiO_2$, on which the perovskite is directly deposited onto FTO, as displayed in Fig S1D.

For $MAPbI_3$ perovskite ($P_1$), an equimolar precursor solution containing 1.42M $PbI_2$ (TCI Chemicals), 1.42 M methyl ammonium iodide (MAI, Dyesol) and 1.42M dimethyl sulfoxide (DMSO, anhydrous) in N,N-dimethylformamide (DMF, anhydrous) was spun-coated onto the mesoporous $TiO_2$ substrates. Few seconds after the deposition, a washing step was performed by dripping 500µL of anhydrous diethylether which produces the bleaching of the films. After that, the films were heated at 100ºC for 5min to form the $MAPbI_3$ film.

In the case of $(MA_{0.17}FA_{0.83})Pb(Br_{0.17}I_{0.83})_3$, also known as $P_2$, a solution made of 1.10M $PbI_2$ (TCI Chemicals), 0.20M $PbBr_2$ (Alfa Aesar), 1.00M formamidinium iodide (FAI, dyesol) and 0.20M methyl ammonium bromide (MABr, Dyesol) was prepared in a solvent mixture of DMSO:DMF (4:1 in v/v). After vigorous stirring, precursor solution was deposited by spin coating using 2 plateaus, the former one at 2000rpm to deposit the precursor solution in dynamic mode and the latter one at 6000rpm to drip 100µL of chlorobenzene. Then, the samples were heated at 100ºC during 30min.

When inorganic monovalent cations such as $Cs^+$ ($P_3$) or $Rb^+$ ($P_4$) were included in the perovskite formulation, the appropriate quantity of $Cs^+$ and/or $Rb^+$ was added to $P_2$ solution from stock solutions made of CsI or RbI (1.5M in DMSO) respectively to achieve the final desired $Cs_xRb_y(MA_{0.166}FA_{0.833})_{1-x-y}Pb(Br_{0.166}I_{0.833})_3$ perovskite as reported in literature[1]. In these cases, deposition and heating procedures were identical to $P_2$ perovskite.

In the case of complete solar cells for photovoltaic characterization, a Spiro-OMeTAD film (2,2',7,7'-tetrakis(N,N-di-p-methoxyphenyamine)-9,9-spirobifluorene, Solarpur®, Merck) was employed as hole selective layer. 35µL of solution were deposited by spin coating in dynamic mode at 3000rpm for 20-25s. This solution was made by dissolving 110mg of Spiro-OMeTAD in 1mL of chlorobenzene; tris(2-(1H-pyrazol-1-yl)-4-tert-butylpyrydine)cobalt(III) bis(trifluoromethylsulphonyl)imide (FK209, Dyesol), lithium bis(trifluoromethylsulphonyl)imide (LiTFSI, Sigma Aldrich) and 4-tert-butylpyridine (*t*-BP, Sigma Aldrich 96%) were added to the solution as additives in relative molar concentrations of 5%, 50% and 330% respectively regarding Spiro-OMeTAD. Perovskite and Spiro-OMeTAD solutions were prepared and deposited under controlled ambient conditions inside a $N_2$ glove box. Finally, 100nm of gold were thermally evaporated under vacuum to obtain the cathode contact of the solar cells. One of these solar cells is displayed in Fig S1B.

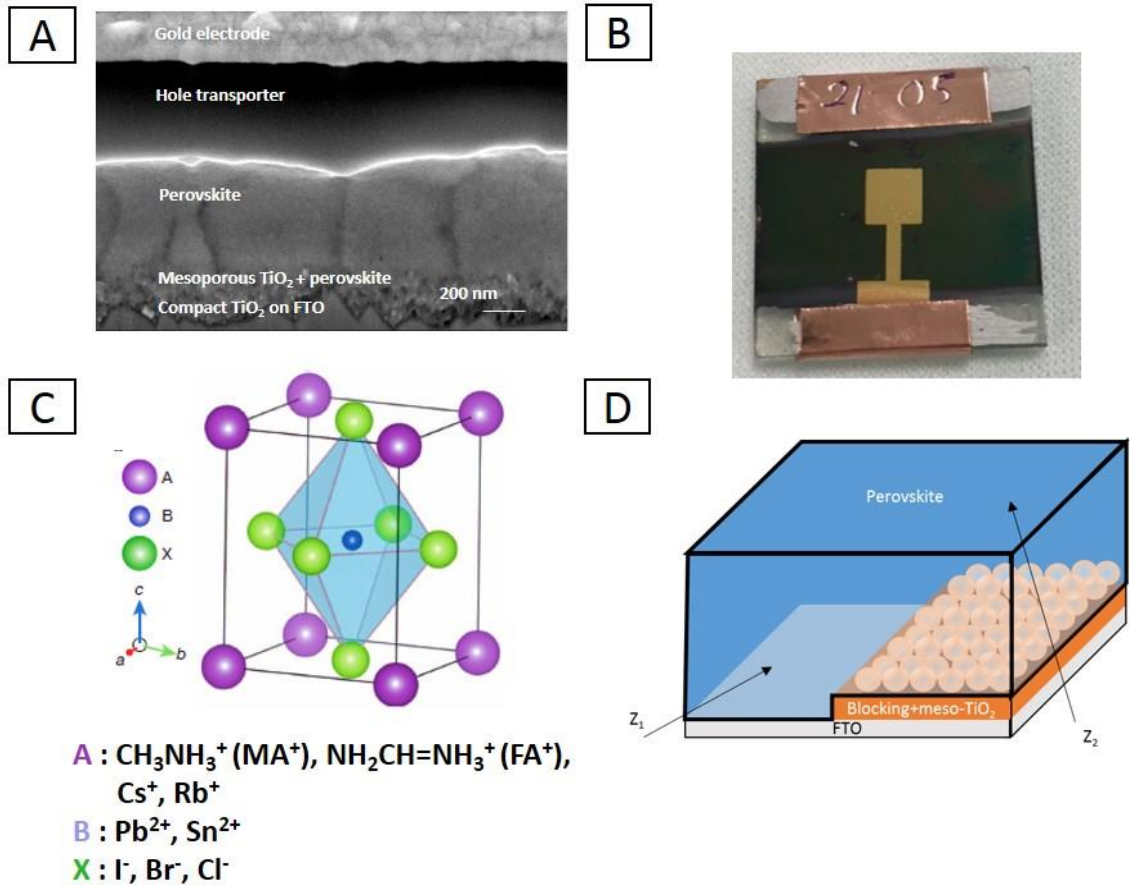

*Figure S1 A- Focused Ion Beam (FIB) image of an (MA$_{0.83}$FA$_{0.17}$Pb(I$_{0.83}$Br$_{0.17}$)$_3$ perovskite solar cell, highlighting the layer structure of the sample measured by I(V) scanning. B- Image of a full device used for photovoltaic characterization (the golden square is 4x4mm) (C) Cubic crystal structure of hybrid organic-inorganic halide perovskites ABX$_3$ ; the cation A could be methylammonium, formamidinium, cesium, rubidium or a mixed of them ; the cation B could be lead (2+) ion or thin (2+) ion and the anion X which is an halogen could be iodide, bromide or chloride anions (D) Sketch showing a lateral view of sample architecture, with observation zones Z$_1$ (perovskite/FTO) and Z$_2$ (perovskite/meso-TiO/blocking-TiO/perovskite).*

## I-V Characteristics Measurement

For *J-V* curves, a solar simulator Oriel was used as light source. For quantum efficiency measurements, an Oriel IQE200 system was utilized to record the performance. In order to allow light biasing, a light bias amplifier (Newport Oriel 70714) was employed when necessary. The active area of the devices was 0.16cm² (4x4mm), while for *J-V* and EQE measurements the active area was fixed to 0.09cm² (3x3mm) with a black metal mask.

## About $n_{T0}$ calculation

Three rate equations model the whole transport in Stranks' recombination model:

$$\frac{\partial n}{\partial t} = -R_{eh}np - R_{pop}n(N_T - n_T) \quad (ES1)$$

$$\frac{\partial n_T}{\partial t} = R_{pop}n(N_T - n_T) - R_{dep}n_T p \quad (ES2)$$

$$\frac{\partial p}{\partial t} = -R_{eh}np - R_{dep}n_T p \quad (ES3)$$

As the typical evolution time for the trapped population is slower than the TRPL decay, it can be interesting to express the mean trapped electron concentration between two pulses:

$$\overline{n_T} = f_0 \int_{t=0}^{t=1/f_0} n_T(t)\, dt \quad (ES4)$$

Considering that the traps are empty when the laser is turned on, $\overline{n_T}$ is growing after each pulse and always lower than $N_T$. Hence, it converges to a value we call $n_{T,0}$. Here, it should be noted that we model the trapped electron population as homogeneous in the depth of the absorber. This corresponds to a fair assumption, as the inhomogeneous regime lasts approx. 100ns, whereas the Laser pulses hit the sample every 40 μs. [*In a practical way, we numerically determine $n_{T,0}$ by running (ES1), (ES2) & (ES3) for each set of fitting parameters, before launching the optimization algorithm.*]

### About $L_{1sun}$ and $\tau_{1sun}$ calculation

As explained in the Model part, equations (ES-1-3) are solved under steady state-illumination, leading to results displayed in Table S1. Interestingly, $n_{T1sun}$ and $n_{1sun}$ are significantly impacted by $N_T$ variations.

| Sample | P$_1$ | P$_2$ | P$_3$ | P$_4$ |
|---|---|---|---|---|
| $n_{1sun}$ [/cm³] | 2.03x10$^{16}$ | 8.1x10$^{15}$ | 7.82x10$^{15}$ | 1.2x10$^{16}$ |
| $n_{T1sun}$ [/cm³] | 1.98x10$^{16}$ | 5.6x10$^{14}$ | 4.03x10$^{14}$ | 7.26x10$^{15}$ |
| $\tau_{1sun}$ [ns] | 43 | 886 | 1009 | 440 |
| $L_{1sun}$ [nm] | 58 | 660 | 778 | 369 |

*Table S1 Parameters used for $L_{1sun}$ and $\tau_{1sun}$ calculations*

### Bi-exponential fitting

Some TRPL transients were fitted with a basic bi-exponential model, leading to lifetimes which cannot be associated to a recombination pathway, and hence lack physical meaning.

| | Sample P$_2$ MA$_{0.83}$FA$_{0.17}$...Pb(I$_{0.83}$Br$_{0.17}$)$_3$ | | | Sample P$_3$ (MA$_{0.83}$FA$_{0.17}$)$_{0.95}$...Cs$_{0.05}$...Pb(I$_{0.83}$Br$_{0.17}$)$_3$ | | |
|---|---|---|---|---|---|---|
| | $\Phi_{0,1}$ | $\Phi_{0,2}$ | $\Phi_{0,3}$ | $\Phi_{0,1}$ | $\Phi_{0,2}$ | $\Phi_{0,3}$ |
| $\tau_1$ [ns] on FTO | 30 | 26 | 14 | 16 | 16 | 12 |
| $\tau_1$ [ns] on mp-TiO$_2$ | 24 | 22 | 13 | 17 | 16 | 11 |
| $\tau_2$ [ns] on FTO | 500 | 271 | 110 | 500 | 250 | 80 |
| $\tau_2$ [ns] on mp-TiO$_2$ | 287 | 181 | 83 | 130 | 92 | 40 |

*Table S2 Lifetimes extracted from a bi-exponential fit of TRPL transients acquired on P$_2$ and P$_3$ at multiple excitation fluxes.*

## Photoluminescence Spectra and $\alpha_{PL}$ calculation

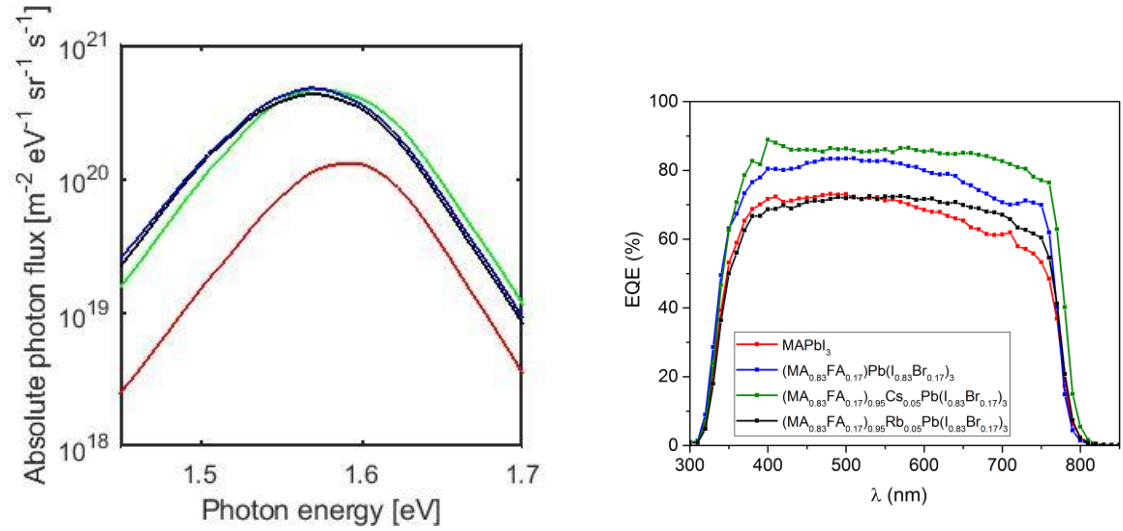

*Figure S2 (left) Absolute PL spectrum for the considered perovskite samples, optically excited at $\lambda$ = 532 nm at $\Phi_0$ = 5.5 x10$^3$ W·m$^{-2}$. No additional peak indicating secondary emitting phase could be observed.(right) .External Quantum Efficiency measurements on full devices.*

The $\alpha_{PL}$ coefficient is the average absorption coefficient for the photoluminescence signal, which can be determined from the $\alpha(E)$ extracted from external quantum efficiency measurement. For this purpose, we assume first that the normalized EQE reflects the internal quantum efficiency (IQE) of the device and then that IQE(E) can be approximated by the internal absorbance of the device A(E). As A(E) = (1-exp(-α(E)*z$_0$), it follows that the absorption coefficient is : α(E)~-1/z$_0$*ln(1-IQE(E)). This estimation is limited around the bandgap of the absorber. Then, $\alpha(E)$ can be convolved with the PL spectrum $\Phi_{PL}(E)$ displayed in Fig S1, with the following formula

$$\alpha_{PL} = \frac{\int \alpha(E)\Phi_{PL}(E)dE}{\int \Phi_{PL}(E)dE} (ES7)$$

Calculations lead to 2.5±1 x 10$^3$ /cm for P$_1$ P$_2$ and P$_4$ and to 5±1 x 10$^3$ /cm for P$_3$.

## In-depth diffusion of photo-generated charge carriers

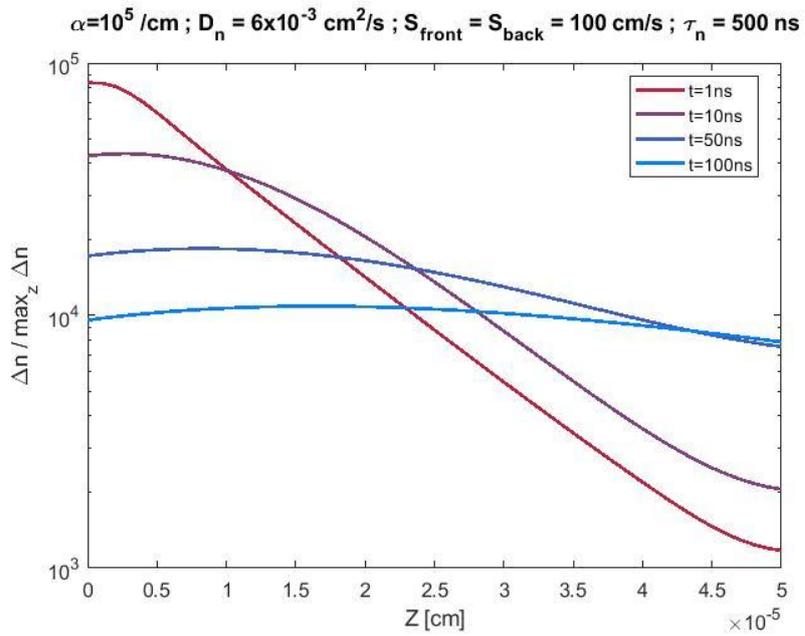

*Figure S3 In-depth concentration profile for electrons (or holes) inside a perovskite layer, following an optical pulsed excitation at $\lambda$ = 532 nm. This calculation is realized with a basic diffusion-recombination model, using the optoelectronic properties determined throughout our TRPL study.*

## TRPL Transients (experimental and fitted) + Reconstruction error for P$_1$ P$_3$ and P$_4$

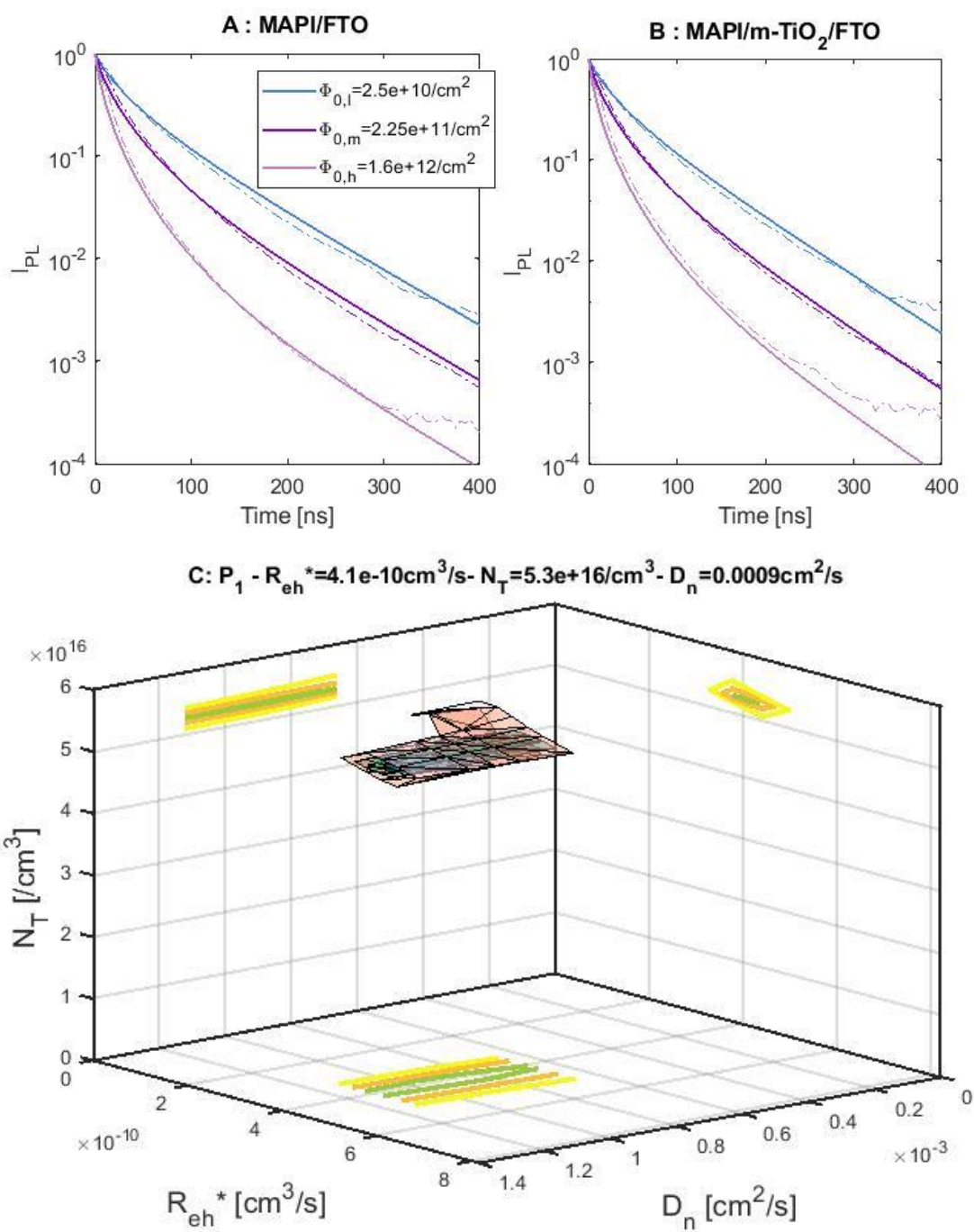

Figure S4 Experimental (dotted lines) and numerically fitted (plain lines) TRPL transients for a MAPbI$_3$ layer deposited on FTO (A) or on mesoporous-TiO$_2$ (B). Various orders of magnitude of the fluence level $\Phi_0$ have been tested. (C) Iso-surfaces showing a set of parameters ($R_{eh}$* $D_n$ $N_T$) leading to a reconstruction error not higher than 10 % (green), 20% (orange), 50% (yellow) of the optimal reconstruction error. The green-orange one is used for confidence intervals.

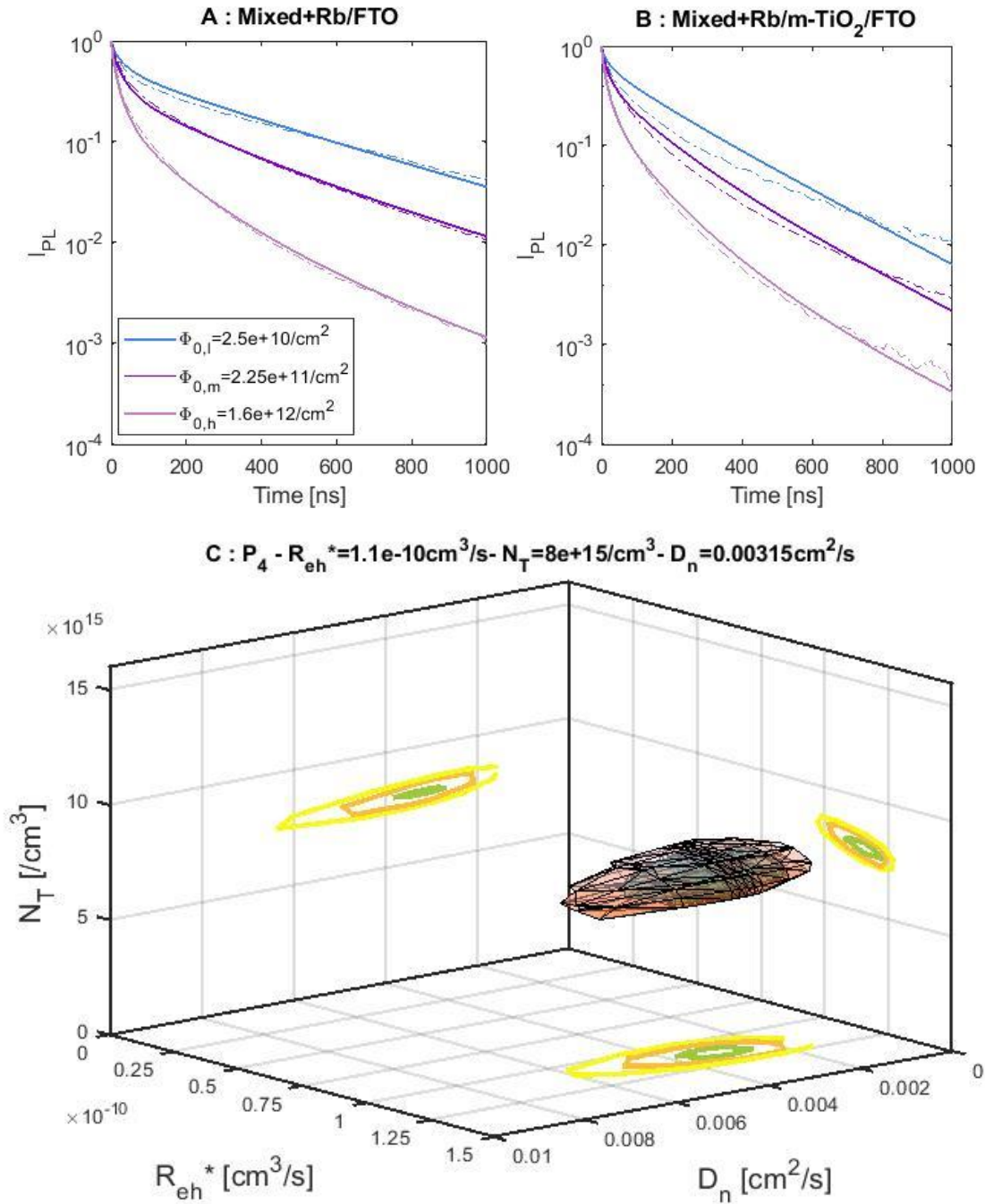

*Figure S5 Experimental (dotted lines) and numerically fitted (plain lines) TRPL transients for a mixed cation perovskite containing 5% Rubidium $(MA_{0.83}FA_{0.17})_{0.95}Rb_{0.05}Pb(I_{0.83}Br_{0.17})_3$ layer deposited on FTO (A) or on mesoporous-TiO$_2$ (B). Various orders of magnitude of the fluence level $\Phi_0$ have been tested. (C) Iso-surfaces showing a set of parameters ($R_{eh}$* $D_n$ $N_T$) leading to a reconstruction error not higher than 10 % (green), 20% (orange), 50% (yellow) of the optimal reconstruction error. The green-orange one is used for confidence intervals.*

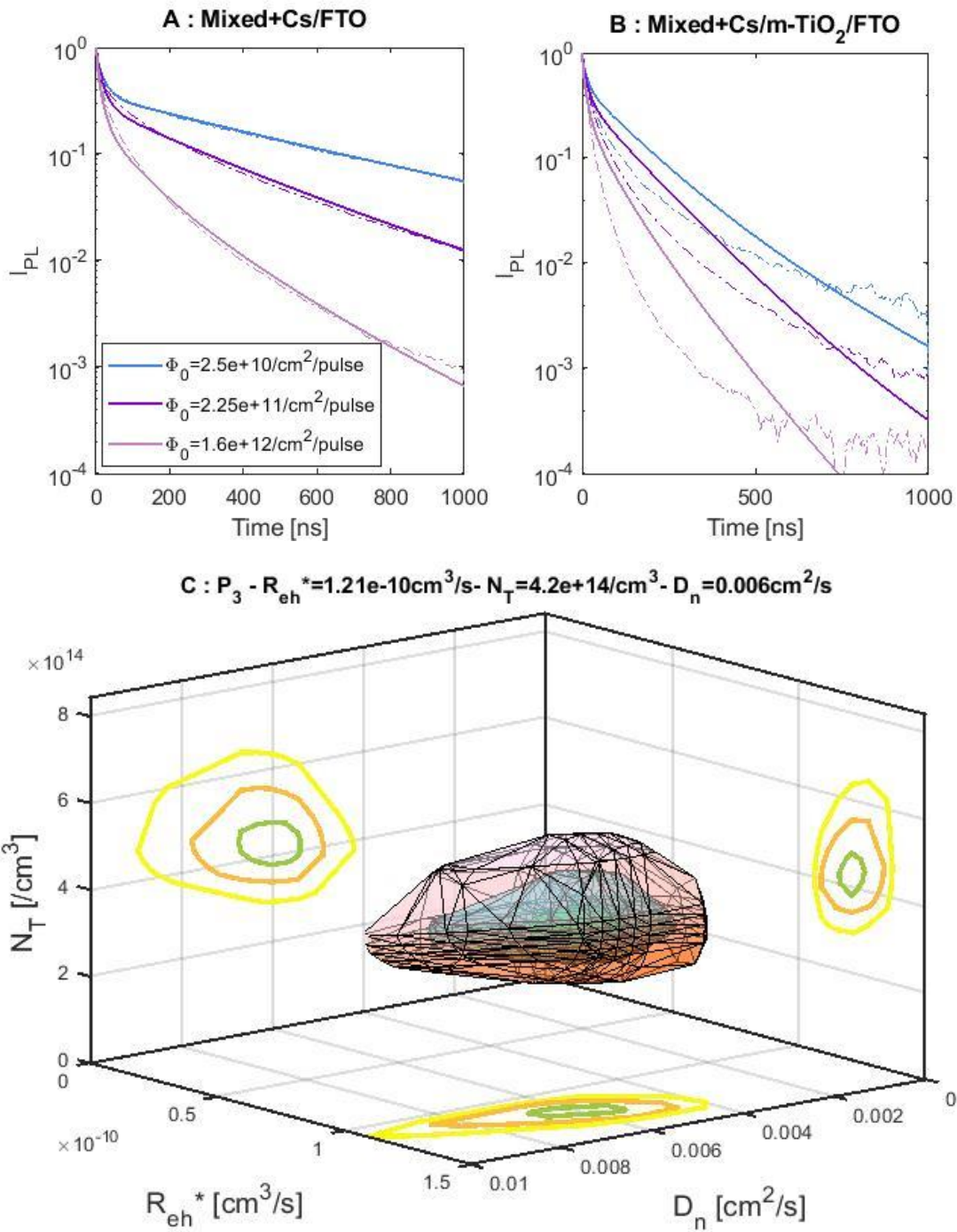

*Figure S6 Experimental (dotted lines) and numerically fitted (plain lines) TRPL transients for a mixed cation perovskite containing 5% Cesium (MA$_{0.83}$FA$_{0.17}$)$_{0.95}$Cs$_{0.05}$Pb(I$_{0.83}$Br$_{0.17}$)$_3$ layer deposited on FTO (A) or on mesoporous-TiO$_2$ (B) [NOT CONVERGED FIT FOR B].Various orders of magnitude of the fluence level $\Phi_0$ have been tested (C) Iso-surfaces showing a set of parameters (R$_{eh}$* D$_n$ N$_T$) leading to a reconstruction error not higher than 10 % (green), 20% (orange), 50% (yellow) of the optimal reconstruction error. The green-orange one is used for confidence intervals.*

## Contribution of each recombination pathway in the charge carrier kinetics

In order to attribute early and later time PL kinetics to different terms in the rate equation, it is interesting to evaluate the impact of each recombination pathway on the charge carrier kinetics. For this purpose, we integrated over the depth each total recombination current with the following formulas.

$$R_{rad}(t) = \int_0^{z_0} R_{eh}^* n(t,z)[n_{T0} + n(t,z)]dz \quad (ES8)$$

$$R_{pop}(t) = \int_0^{z_0} R_{pop} n(t,z)[N_T - n_{T0}]dz \quad (ES9)$$

$$R_{int}(t) = S_{front} \times n(t,z=0) \quad (ES10)$$

For the three excitation fluences, the temporal evolution of each recombination pathway as well as a summary of the main recombination mechanisms at short and long times are displayed below.

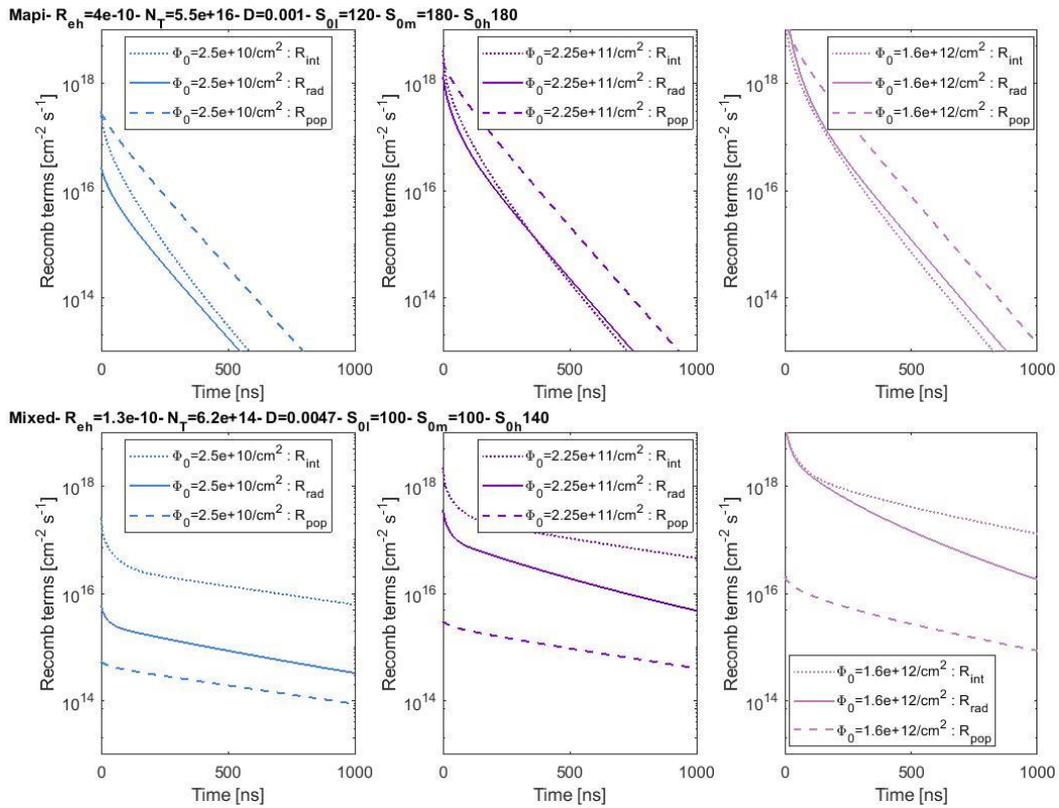

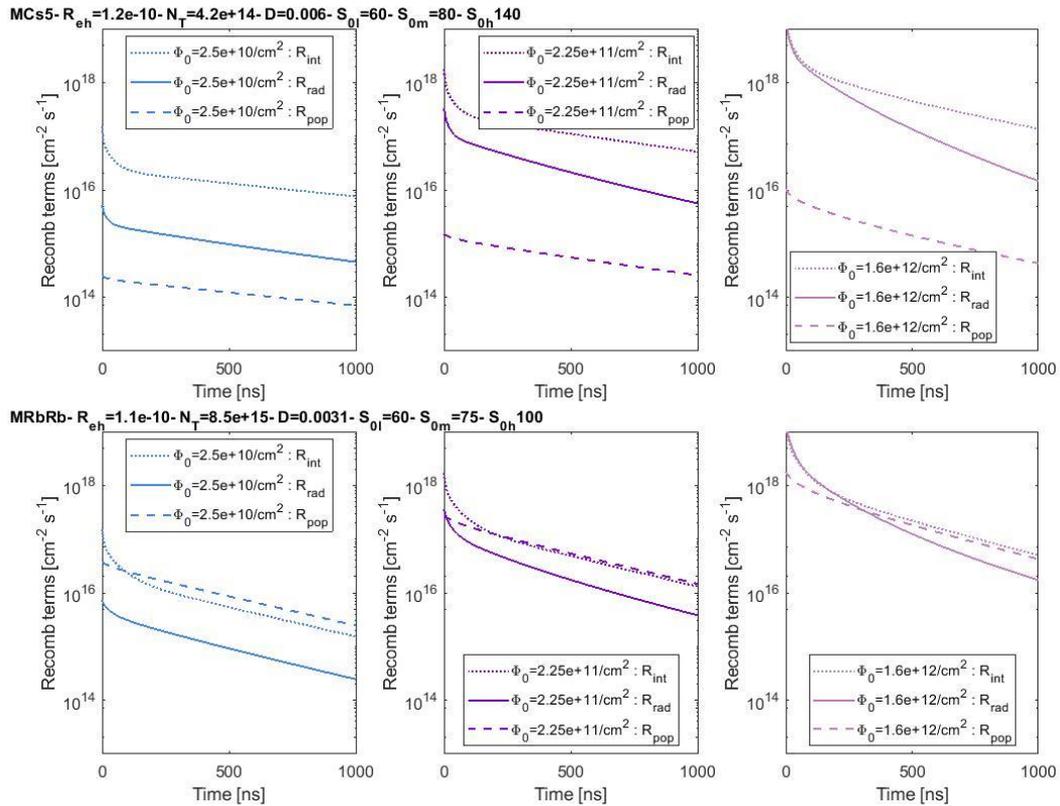

*Figure S7 Temporal evolution of recombination terms simulated for various perovskite compositions, at the same excitations powers employed during TRPL decay.*

|  | MAPbI$_3$ | | Perovskite P$_2$ | | Perovskite P$_3$ | | Perovskite P$_4$ | |
|---|---|---|---|---|---|---|---|---|
|  | Short times | Long times | Short times | Long times | Short times | Long times | Short times | Long times |
| $\Phi_{0,1}$ | Interface + Traps | Trap-assisted | Interface (diffusion away from) | Interface | Interface (diffusion away from) | Interface | Interface (diffusion away from) | Trap-assisted |
| $\Phi_{0,2}$ | Interface + Traps | Trap-assisted | Interface (diffusion away from) | Interface | Interface (diffusion away from) | Interface | Interface (diffusion away from) | Interface + Traps |
| $\Phi_{0,3}$ | Traps + Direct | Trap-assisted | Interface + Direct | Interface | Interface + Direct | Interface | Interface + Direct | Interface + Traps |

*Table S3 Major recombination pathways at short times (t = 0.. 20 ns) and long times (t>100 ns), calculated for each perovskite composition studied here.*

## Control experiment for perovskite P$_3$ deposited on glass substrate

To verify our approximation of neutral interface at FTO/perovskite, we realized a control experiment for a sample with composition P$_3$, deposited on glass. TRPL decays observed for perovskite on glass and on FTO are similar for the different excitation intensities, even if the latter is slightly accelerated at higher excitation flux. However, the difference between FTO and glass decays is smaller than the one between FTO and mp-TiO$_2$. This is illustrated in Figure S8. Decays have been fitted and corresponding parameters are indicated in Table S4, along with experimental and numerical transients, as well as the cube showing the reconstruction errors. The fitting constants converge to similar values for both acquisitions, except for the $R_{eh}*$ coefficient, which is 40% higher for the fit on FTO. As indicated in the discussion part, TRPL cannot effectively distinguish between back surface and bulk recombination[2]. For this purpose, TRPL should be acquired by illuminating on FTO side and on glass side, and transients should be fitted together.[3]

In any case, the observed difference between samples deposited on glass and on FTO do not suggest us to integrate a significant $S_{back}$ at perovskite/FTO interface. This justifies the fact that $S_{FTO} = 0$ is taken as a reference for $S_{back}$, to which mesoporous-TiO2 back interface can then be compared, when decays are fitted on this architecture.

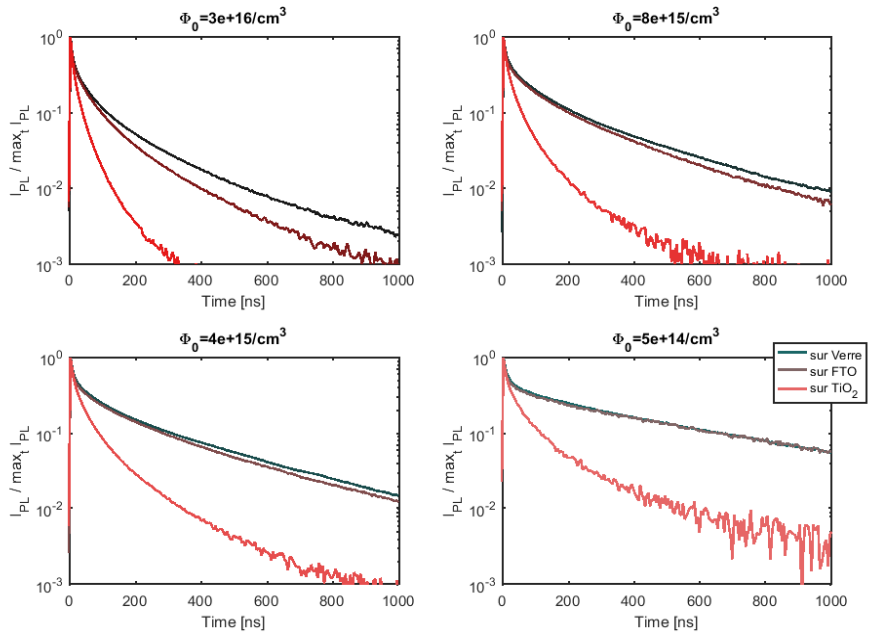

Figure S8 TRPL decays observed for the composition $P_3$, deposited on glass (green curve), on FTO (brown curve) and on mp-TiO$_2$ (red curve). Various excitations intensities have been used. In this figure only, $\Phi_0$ designated the electron concentration in the perovskite thin film.

| Mixed + 5%Cs | $R_{eh}*$ | $R_{eh}=R_{eh}**g_{corr}$ | $N_T$ | $S_{front}$ | $S_{back}$ | $D_n$ |
|---|---|---|---|---|---|---|
| /glass | 7±1 x10$^{-11}$ | 5.6x10$^{-10}$ [using ref[4]] | 1.2±1 x10$^{14}$ | 60/80/140 | Assumed 0 | 4.2±1.5 x10$^{-3}$ |
| /FTO/glass | 1.2±0.1 x10$^{-10}$ | ?? | 4.2±2 x10$^{14}$ | 60/80/140 | Assumed 0 | 6±1.5 x10$^{-3}$ |

Table S4 Fitting parameters for perovskite deposited on glass and on FTO.